\documentstyle[prl,aps,preprint]{revtex}
\begin{document}
\setcounter{page}{0}
\def\footnoterule{\kern-3pt \hrule width\hsize \kern3pt}
\tighten
\title{   Hidden conformal symmetry of a massive scalar field in
\\AdS$_2$
 \thanks
{This work is supported in part by funds provided by the U.S.
Department of Energy (D.O.E.) under cooperative 
research agreement \#DF-FC02-94ER40818.
}}
\author{J. Cruz
\footnote{Email addresses:
 {\tt cruz@ctpa03.mit.edu \\ cruz@lie.uv.es}}}
\address{Center for Theoretical Physics \\
Laboratory for Nuclear Science \\
and Department of Physics \\
Massachusetts Institute of Technology \\
Cambridge, Massachusetts 02139 \\
and \\
Departamento de Fisica Teorica and IFIC\\
Centro Mixto Universidad de Valencia-CSIC\\
Facultad de Fisica,
Universidad de Valencia\\
 Burjassot-46100, Valencia, Spain
{~}}
\date{MIT-CTP-2752,~ hep-th/9806145. {~~~~~} June 1998}
\maketitle
\thispagestyle{empty}
\begin{abstract}
We show that a massive scalar field on a 2-dimensional Anti-De Sitter
space (AdS$_2$) can be mapped by means 
of a time-dependent canonical transformation
into a
 massless free field
 theory when the mass
 square of the field is equal to minus  the curvature of the background metric.
 We also provide the (hidden) conformal symmetry of the massive
scalar field.
\end{abstract} 
\bigskip
PACS number(s): 04.60.Kz, 04.60.Ds
\newline
\vspace*{\fill}
\begin{center}
Submitted to: {\it Phys. Lett. B}
\end{center}
The Liouville action 
\begin{equation}
S_L=-{1\over2}\int d^2 x \left[ \partial_{\mu}\phi\partial^{\mu}\phi+
{\lambda^2\over \beta^2}e^{\beta\phi}\right]\>,\label{liouville}
\end{equation}
defined over the 2-dimensional Minkowski space, is not invariant under
conformal transformations when the field $\phi$ transforms in the 
usual way
\begin{equation}
\delta\phi=f^{\mu}\partial_{\mu}\phi\>,\label{o}
\end{equation}
where $f^{\mu}$ is a conformal Killing vector (i.e, the generator
 of the transformation). 
This  transformation rule leaves invariant, up to a boundary term,
the free field theory ($\lambda=0$).
To achieve the symmetry transformation in the Liouville model
the field $\phi$ must transform as follows
\begin{equation}
\delta\phi=f^{\mu}\partial_{\mu}\phi+{1\over\beta}
\partial_{\mu}f^{\mu}\>.\label{i}
\end{equation}
The Liouville field ${\phi}$ can be converted
 by a canonical transformation into
a free field $\psi$ \cite{Liouville}. 
The aim of this letter is to prove that  a similar construction can be 
made if the auto-interacting term in (\ref{liouville}) is replaced by a mass
term $\lambda^2\phi^2$ and the scalar field is minimally coupled to
a  AdS$_2$ backgroud metric \footnote{
Actually we will work on the universal covering space CAdS$_2$ to avoid
the problem of the existence of closed time-like
 curves, although we will use the same notation for simplicity} 
  with curvature
$R=-\lambda^2$.
However in this case the transformation mapping the theory into a
(massless) free
field theory contains an explicit dependence on time.
 This implies
that  the free field dynamics is generated by a Hamiltonian which
differs from the initial one by a total derivative term in
the extended phase-space.  
As a consequence of our construction
the massive field theory is invariant under conformal transformations
when the massive field transforms
 in an involved  way which is implicitly defined
by the  equations of the canonical transformation.
    
Let us consider the action of a 2-dimensional massive scalar field $\phi$
minimally coupled to 
a background metric $g_{\mu\nu}$
\begin{equation}
S=-{1\over2}\int d^2x\sqrt{-g}\left[\left(\nabla\phi\right)^2+\lambda^2
\phi^2\right]\label{massive}
\>.
\end{equation}
If we express the metric in the conformal gauge
$\left(ds^2=-e^{2\rho}dx^+dx^-\right)$, then we can write
\begin{eqnarray}
S&=&\int dx^+dx^-\left[\partial_+\phi\partial_-\phi-{\lambda^2\over4}e^{2\rho}
\phi^2\right]\nonumber\\&&
={1\over2}\int dtdx \left[{\dot \phi}^2-\phi^{\prime 2}
-\lambda^2\phi^2 e^{2\rho}\right]
\>.\label{confgauge}
\end{eqnarray}
We denote with dot derivation with respect to the time coordinate
$t={1\over2}(x^++x^-)$ and with prime derivation with respect to the spatial
coordinate $x={1\over2}(x^+-x^-)$.
 The hamiltonian form of the action becomes
\begin{equation}
S=\int dtdx \left[ \dot\phi\pi_{\phi}-\cal H \right]
\>,
\end{equation}
where the hamiltonian density ${\cal H}$ is
\begin{equation}
{\cal H }={1\over2}\left[\pi_{\phi}^2+\left(\phi^{\prime}\right)^2
\right]
+{\lambda^2\over2}\phi^2 e^{2\rho}
\>.\label{hache}
\end{equation}
By choosing the background metric to have constant curvature
  $-\lambda^2$ we obtain
 \begin{equation}
 R=8e^{-2\rho}\partial_+\partial_-\rho=-\lambda^2\>,\label{R}
 \end{equation}
 which is the familiar 
 Liouville equation
with solution
\begin{equation}
\rho={1\over2}\log{ \partial_+A_+\partial_-A_-
\over \left(1+{\lambda^2\over8}A_+A_-\right)^2}\label{ads1}
\>,
\end{equation}
where $A_{\pm}(x^{\pm})$ are arbitrary
 chiral functions.
We fix the gauge by setting $A_{\pm}=x^{\pm}$.
The metric with conformal factor (\ref{ads1}) is the AdS$_2$
metric with negative 
 curvature $-\lambda^2$ in conformal coordinates $x^{\pm}$.
Now the equation of motion for the field $\phi$ is
\begin{equation}
\partial_{+}\partial_-\phi=-{\lambda^2\over4}e^{2\rho}
\phi=-{\lambda^2\over4}{\phi \over 
\left(1+{\lambda^2\over8}x^+x^-\right)^2}\label{eq7}
\>.
\end{equation}
It is easy to check that an arbitrary solution to this equation
satisfies
\begin{equation}
\partial_-(e^{2\rho}\partial_+(e^{-2\rho}\partial_+\phi))=0\>,
\label{auxi}
\end{equation}
\begin{equation}
\partial_+(e^{2\rho}\partial_-(e^{2\rho}\partial_-\phi))=0\>.
\label{auxii}
\end{equation}
Let us consider the equation (\ref{auxi}). The general solution is
\begin{equation}
\phi=\int^+ \left[ e^{2\rho}\int^+\left(e^{-2\rho}b_+\right)\right]
+a_-\partial_-\rho
+b_-\>,\label{ins}
\end{equation} 
 where  $a_-(x^-), b_{\pm}(x^{\pm})$ are arbitrary chiral functions.
This is not, in general, a solution to the equation (\ref{eq7}).
However if we set $b_+=0$ and insert (\ref{ins}) into (\ref{eq7})
 we can see that 
it is a solution provided 
\begin{equation}
b_-={1\over 2} \partial_-a_-\>,
\end{equation}
and then
\begin{equation}
\phi={1\over2}\partial_-a_--{\lambda^2\over8}{x^+a_-\over
1+{\lambda^2\over8}x^+x^-}\>.
\end{equation}
A similar argument using (\ref{auxii}) shows that 
\begin{equation}
\phi=-{1\over2}\partial_+a_++{\lambda^2\over8}{x^-a_+\over 1+
{\lambda^2\over8}x^+x^-}\>,
\end{equation}
is also a solution to the equation (\ref{eq7}).
Because the equation is linear the general solution is
 therefore
\begin{equation}
\phi=-{1\over2}\left(\partial_+a_+-\partial_-a_-\right)+{\lambda^2\over8}
{a_+x^--a_-x^+\over 1+{\lambda^2\over8}x^+x^-}\>,\label{sol}
\end{equation}
 with $a_{\pm}(x^{\pm})$ arbitrary chiral functions.
It is fairly well-known that the form of the general solution (\ref{ads1})
to the Liouville equation induces a canonical transformation between the
Liouville theory and a massless free field theory \cite{Liouville}.
 Mimicking the same procedure
we can construct from (\ref{sol}) the following 
phase-space transformation 
from the variables $\phi,\pi_\phi$ to the new ones $a_+,a_-$
(which are no longer chiral functions)
\begin{equation}
\phi=-{1\over2}\left(a_+^\prime+a_-^\prime\right)+{\lambda^2\over8}
{a_+x^--a_-x^+\over 1+{\lambda^2\over8}x^+x^-}\>,\label{phi}
\end{equation}  
\begin{eqnarray}
\pi_{\phi}&=&-{1\over2}(a_+^{\prime\prime}-a_-^{\prime\prime}
)+{\lambda^2\over8}{a_+^\prime x^-
+a_-^\prime x^+\over  1+{\lambda^2\over8}x^+x^-}\nonumber\\&&
+{\lambda^2\over8}{a_+-a_-\over 1+{\lambda^2\over8}x^+x^-}
-{\lambda^4\over64}{(x^++x^-)(a_+x^--a_-x^+)\over
(1+{\lambda^2\over8}x^+x^-)^2}
\>.\label{piphi}
\end{eqnarray}
Defining 
\begin{equation}
\psi=-{1\over2}(a_+^{\prime}+a_-^{\prime}),\qquad
\pi_{\psi}=-{1\over2}(a_+^{\prime\prime}-a_-^{\prime\prime})
\>,\label{psipipsi}
\end{equation}
the transformation $\phi,\pi_{\phi}\rightarrow\psi,\pi_{\psi}$ is canonical
up to boundary terms.
This can be seen by evaluating the symplectic form
\begin{equation}
\omega= \int dx\ \delta \phi \wedge\delta\pi_{\phi}
\>.
\end{equation}
Substituting the transformation (\ref{phi}),(\ref{piphi}),(\ref{psipipsi})
 and after some computation it can be written as
(from now on we shall omit the symbol of exterior product)
\begin{eqnarray}
\omega&=& -{1\over4}\int dx \ \left[\delta a_+^\prime
\delta a_+^{\prime\prime}-\delta a_-^{\prime}
\delta a_-^{\prime\prime }\right]+\int d\left[ {1\over4}
 \delta a_+^\prime\delta a_-^\prime
\right.\nonumber\\&&\left.
+{\lambda^2\over16}{1\over 1+
{\lambda^2\over8}x^+x^-}\left(x^-\delta a_+\delta a_+^\prime
-x^-\delta a_+\delta a_-^\prime-x^
+\delta a_-\delta a_+^\prime +x^+\delta a_-\delta a_-^\prime\right)\right]
\nonumber\\&&=
\int dx\ \delta\psi\delta\pi_{\psi}+\omega_b
\>,
\end{eqnarray}
where $\omega_b$ is a boundary term.
At this stage we must recall that  AdS$_2$  is not a  globally hyperbolic
space-time, that is, there is no a well-defined Cauchy problem.
In order to make the Cauchy problem well-defined we must specify certain
boundary conditions of the fields at spatial infinity \cite{Avis}.
These conditions ensure that quantities like the energy of the solutions,
the Klein-Gordon product of two solutions to the equation of motion
or the symplectic form are conserved (independents of time).
Specifically, if we impose that asymptotically   $a_{\pm}$
behaves like
\begin{equation}
a_{\pm}\sim (x^{\pm})^{ 1-\epsilon}\qquad  0<\epsilon<<1 \qquad
|x^{\pm}| \rightarrow\infty\>,\label{bound}
\end{equation}
then all the fields fall off to zero  rapidly enough to make all
 the expressions well-defined.
With these conditions 
the bulk part of $\omega$ is a convergent expression and on the other hand
the boundary contribution $\omega_b$ vanishes and then the theory is truly 
canonically equivalent to the theory of a field $\psi$ which
clearly satisfies the (massless) free field equations
\begin{eqnarray}
\dot\psi=\pi_{\psi},\qquad \dot \pi_{\psi}=\psi^{\prime\prime}
\>.
\end{eqnarray}
Because the transformation is time-dependent these equations are not
derived from the hamiltonian density (\ref{hache}).
To see this explicitly
let us consider the following 1-form in the extended phase-space
 with coordinates
$(\phi(x),\pi_{\phi}(x),t)$
\begin{equation}
\Omega=\int dx \left[\pi_{\phi}(x)\delta \phi(x)\right]-Hdt
\>.
\end{equation}
Under a canonical transformation to the variables $(\psi(x),\pi_{\psi}(x),t)$
$\Omega$ transforms in the following way
\begin{equation}
\Omega=\int
 dx\left[\pi_{\psi}(x)\delta \psi(x)\right]-\tilde H dt+\delta K
\>,
\end{equation}
for some functionals $\tilde H$ and $K$ of the extended
 phase-space coordinates.
Then the hamiltonian equations of motion in terms
 of $\psi,\pi_{\psi}$ are derived from the new
hamiltonian $\tilde H$
\begin{equation}
\dot \psi = \{\psi,\tilde H\}_{P.B}={\delta \tilde H\over\delta\pi_{\psi}}
\>,
\end{equation}
\begin{equation}
\dot \pi_{\psi} = \{\pi_{\psi},\tilde H\}_{P.B}=-{\delta \tilde H\over\delta\psi}
\>.
\end{equation}
In the present case this is satisfied with
\begin{equation}
\tilde H=
{1\over2}\int  dx\left[\left(\psi^{\prime}\right)^ 2
+\pi_{\psi}^2\right]
\>,\label{tilde}
\end{equation}
and
\begin{eqnarray}
K&=&\int dx \left[-{\lambda^2
\over16}{(x^++x^-)a_+^\prime a_-^\prime\over
1+{\lambda^2\over8}x^+x^-}-{\lambda^2\over16}{a_+a_+^\prime-a_-
a_-^\prime\over 1+{\lambda^2\over8}x^+x^-}\right.\nonumber\\&&
+{\lambda^4\over128}{(x^++x^-)(a_+^\prime+a_-^\prime)(a_+x^--a_-x^+)
\over \left(1+{\lambda^2\over8}x^+x^-\right)^2}\nonumber\\&&
+{\lambda^4\over128}{(a_+x^--a_-x^+)(
a_+-a_-)\over \left(1+{\lambda^2\over8}x^+x^-\right)^2}
\nonumber\\    
 &&\left.
-{\lambda^6\over1024}{(x^++x^-)(a_+x^--a_-x^+)^2\over
\left(1+{\lambda^2\over8}x^+x^-\right)^3}\right]
\>.
\end{eqnarray}
$K$  is the time-dependent generating functional of the transformation
given as an implicit
functional of the extended phase-space variables and the boundary contributions
to $\tilde H$ and $K$ vanish by virtue of the boundary conditions
(\ref{bound}).

We have yet said that the massless free field action
\begin{equation}
S=-{1\over2}\int d^2x\ \partial_{\mu}\psi\partial^{\mu}\psi
\>,
\end{equation}
 (and therefore the equation of motion) is invariant, up to
a boundary term,  under the transformation
\begin{equation}
\delta\psi=f^{\mu}\partial_{\mu}\psi
\>.
\end{equation}
if $f^{\mu}$ is a conformal Killing vector.
The transformation (\ref{phi}),(\ref{piphi}),(\ref{psipipsi})
we have just constructed implies that the action
(\ref{massive}) and the equation of motion (\ref{eq7})
 have an implicit hidden symmetry which depends on the coordinates
\begin{equation}
\delta\phi=f^{\mu}\partial_{\mu}\psi+{\lambda^2\over8}{x^-\delta a_+
-x^+\delta a_-\over 1+{\lambda^2\over8}x^+x^-}
\>,\label{sym}
\end{equation}
where the variations $\delta a_{\pm}$ are calculated from (\ref{psipipsi}) with
$\pi_{\psi}=\dot\psi$ and it must be understood that
 all the functions which appear
in (\ref{sym}) are implicit functionals of $\phi$ through
(\ref{phi}),(\ref{piphi}),(\ref{psipipsi}) with $\pi_{\phi}=\dot\phi$.
This rather involved symmetry of the equation of motion (\ref{eq7}) is
the underlying reason which makes possible to find the exact solution
(\ref{sol}).
 
We can 
understand better the presence of a conformal
 symmetry in this theory with the following
argument.
Let us consider the Jackiw-Teitelboim action \cite{Jackiw}
\begin{equation}
S=\int d^2x\sqrt{-g}[R+\lambda^2]\phi\>.\label{JT}
\end{equation}
In this action both the metric and the dilaton
 $\phi$ are dynamical fields, so
the theory is generically covariant.
In conformal gauge, the unconstrained equations of motion are
precisely the Liouville equation (\ref{R}) and the equation for the
massive scalar field (\ref{eq7}).
Both equations are invariant under conformal transformations 
 when $\phi$ behaves as a 
scalar (\ref{o}) and $\rho$ transforms as in
(\ref{i}).
Now if we consider the metric to be a fixed background metric solution
to the Liouville equation (\ref{R}), then the equation (\ref{eq7}) still
maintains the conformal invariance but now $\rho$ remains fixed
and $\phi$ inherits a peculiar transformation law under
conformal transformations which is just the one we have described in this 
letter.

Finally we would like to  point out that when we approach to the
 boundary of the space-time $|x^+x^-|\rightarrow\infty$ the solution
(\ref{sol}) behaves like
\begin{equation}
\phi=-{1\over2}(\partial_+a_+-\partial_-a_- )+{a_+\over x^+}-{a_-\over x^-}
\>,
\end{equation}
that is, like a free field which has a simple relation with $\psi$.
So there is a hidden conformal structure in the theory which becomes clear
when we approach to the boundary.
It would be interesting to see the possible connection of this with
the correspondence between field theories in AdS$_{d+1}$ and
d-dimensional conformal field
theories on the boundary \cite{Maldacena}. 
  These questions will be considered in future research.

\section*{Acknowledgments}
I want to acknowledge  the Generalitat Valenciana for a FPI fellowship.
I want to thank  J. Navarro-Salas for many useful
comments and suggestions and also to K. Bering for discussions
 and  to R. Jackiw for a critical reading of the manuscript.

\end{document}